\documentclass[aps,prd,twocolumn,10pt,superscriptaddress,floatfix,nofootinbib,longbibliography]{revtex4-1}  
\usepackage{color}
\usepackage{slashed}
\usepackage{amsmath}
\usepackage{latexsym}
\usepackage{mathrsfs}
\usepackage{url}
\usepackage{amssymb,amsfonts,amsmath,bm,bbm}   
\usepackage{graphicx}
\usepackage{dcolumn}
\usepackage{bm}
\usepackage[
colorlinks=true,
linkcolor=blue,
breaklinks=true,
urlcolor=blue,
citecolor=blue]{hyperref}
\usepackage{bbold}
\usepackage{epstopdf}
\def\beq{\begin{eqnarray}}
\def\eeq{\end{eqnarray}}
\newcommand{\HNUST}{\affiliation{
Hunan Provincial Key Laboratory of Intelligent Sensors and Advanced Sensor Materials, \\ School of Physics and Electronics, Hunan University of Science and Technology, Xiangtan 411201, China}}

\newcommand{\BU}{\affiliation{
School of Physics, Beihang University, Beijing 102206, China}}

\begin{document}

\preprint{APS/123-QED}

\title{New axion contribution to the two-photon decays of neutral pions
}

\author{Zhen-Yan Lu} \email{luzhenyan@hnust.edu.cn}\HNUST

\author{Yang Huang} 
\email{yanghuang55@outlook.com}
\HNUST

\author{Ji-Gui Cheng}
\email{cheng-jg@hnust.edu.cn}
\HNUST

\author{Qi Lu}\email{luqi0@buaa.edu.cn}\BU

\author{Shu-Peng Wang}\HNUST

\date{\today}

\begin{abstract}
The presence of axions introduces new diagrams at 
one-loop order to the two-photon decays of the neutral pion through axion-pion mixing.  
In this work, we calculate this correction, missing in all current calculations, in the framework of SU(2) chiral perturbation theory. We show that the correction is proportional to the axion-photon coupling and the square of axion mass, which in turn is strongly suppressed by the axion decay constant for the classical space window but may not be negligible for the QCD axion in the MeV or even larger mass range. On the other hand, in combination with the experimental measurement of the decay width of $\pi^0\rightarrow\gamma\gamma$ process, this result rules out the standard QCD axion as an explanation for the possible discrepancy between the chiral perturbation theory prediction and the experimental data.

\end{abstract}

\maketitle

\section{Introduction} \label{INTRO}

The anomalous decay of the neutral pion~\cite{Bernstein-2011bx,Ananthanarayan-2022wsl} into two photons is the most important process in the odd-intrinsic parity sector of Quantum Chromodynamics (QCD)~\cite{Gross-2022hyw} since this is the process by which the quantum anomaly was historically discovered and the color number $N_c=3$ was confirmed.  
This process can be described by the Wess-Zumino-Witten (WZW) Lagrangian, and the leading order (LO) gives 
\begin{eqnarray}
\mathcal{L}_{\pi^0\gamma \gamma}=-
\frac{e^2}{32 \pi^2F_0} \epsilon^{\mu \nu \rho \sigma}\pi^{0} \mathcal{F}_{\mu \nu} \mathcal{F}_{\rho \sigma},
\end{eqnarray}
where $e$ is the elementary charge, $\epsilon^{\mu \nu \rho \sigma}$ is the completely
antisymmetric tensor, $\pi^0$ and $\mathcal{F}_{\mu \nu}$ denote the pion field and the electromagnetic field-strength tensor, respectively. 
As the lightest QCD bound state, 
the neutral pion decays almost exclusively, about 98.8\%,  
into two photons, providing a critical  
process for testing and searching for new physics beyond the Standard Model~\cite{Workman-2022ynf}.

Like the neutral pion motioned above, 
the QCD axion, as a Goldstone boson corresponding to the spontaneous breaking of the global $U(1)_{{PQ}}$ symmetry~\cite{78Weinberg223-226PRL,78Wilczek279-282PRL}, can also decay into two photons through the axion-photon coupling term~\cite{85Kaplan215-226NPB,19Alonso-alvarez.Gavela.ea223-223EPJC}. This hypothetical uncharged particle~\cite{77Peccei.Quinn1440-1443PRL,77Peccei.Quinn1791-1797PRD} was first proposed to solve the strong CP problem. 
Except for photons, the axion can also couple to the other Standard Model (SM) particles, such as nucleons and photons. However, all such couplings are suppressed by the axion decay constant $f_a$, resulting in the invisible axion~\cite{81Dine.Fischler.ea199-202PLB,Zhitnitsky-1980tq,Kim-1979if,Shifman-1979if}, 
which has very weak couplings to the SM particles due to the remarkably large value of $f_a$~\cite{83Preskill.Wise.ea127-132PLB,Antel-2023hkf}.
Since the axion-photon coupling vertex allows for the production of an axion from the interaction of a photon with the magnetic field background, the axion-photon coupling $g_{a\gamma\gamma}$
plays a central role in searching for axions both from laboratory experiments and stellar objects~\cite{03Bradley.Clarke.ea777-817RMP,Jaeckel-2010ni,Ringwald-2012hr,15Graham.Irastorza.ea485-514ARNPS,Sikivie-2020zpn,Chen-2023jki,Semertzidis-2021rxs}. Recently, this coupling has been computed precisely up to the next-to-leading (NLO) order both in SU(2)~\cite{GrillidiCortona-2015jxo} and SU(3)~\cite{Lu-2020rhp} chiral perturbation theory (CHPT). However, the poor knowledge of $f_a$ and 
the anomaly coefficients has so far prevented the determination of the strength of axion-photon coupling and the experimental detection of axions, see Refs.~\cite{Kim-2008hd,16Marsh1-79PR,DiLuzio-2020wdo} for several recent reviews.    

Although many efforts have been focused on the study of the decay width of anomalous two-photon decay of the neutral pion by using various methods~\cite{PrimEx-II-2020jwd,Feng-2012ck,Bijnens-1988kx,Moussallam-1994xp,Goity-2002nn,Borasoy-2003yb,Nasrallah-2002yi,10Bijnens.Kampf220-223NPBS,Kampf-2009tk,Ioffe-2007eg}, to the best of
our knowledge, a careful study of the decay width of this process with the inclusion of the contribution from the axion-pion mixing has never been derived in the literature. 
With the present manuscript, we fill this gap by analyzing the anomalous decay of neutral pion to two-photon in the presence of a background QCD axion field~\cite{ADMX-2023rsk}. We will perform our work in the framework of SU(2) axion CHPT with the inclusion of the WZW Lagrangian accounting for anomalous processes. As an effective field theory of QCD at the low energy scale, the 
prediction of the topological susceptibility up to NLO from SU(2) CHPT is precisely consistent with the lattice simulation: 
\begin{eqnarray}
\chi_t^{1/4}=
\begin{cases}
75.5(5)~\text{MeV},  &\text{CHPT}~\text{\cite{GrillidiCortona-2015jxo}},\cr
75.6(2)~\text{MeV},  &\text{Lattice data}~\text{\cite{16Borsanyi.others69-71N}}.\cr
\end{cases}
\end{eqnarray}
Based on the fact that the SU(2) CHPT can perfectly capture the low-energy properties of the QCD $\theta$-vacuum and the equation of state of strongly interacting QCD matter at low densities~\cite{Adhikari-2019zaj,Adhikari-2020kdn}, 
such as the precise determination of topological susceptibility shown above and prediction for the peak structure of the ratio of energy density to its Stefan-Boltzmann limit~\cite{16Carignano.Mammarella.ea51503-51503PRD}, we expect to have a high precision estimate of the contribution from the axion-pion mixing to the $\pi^0\rightarrow\gamma\gamma$ process. 

We note that in spite of recent progress with the help of lattice QCD, some of the couplings $l_i$ and $h_i$ of CHPT are still poorly known. 
Thus, we emphasize that in this work, we will not go through the determination of coupling strengths of the theory and constraints on the axion parameter space, but instead focus on the estimates of the correction induced by the axion-pion mixing and study its evolution with respect to the axion mass from eV to MeV scale. As we will see, the correction $\delta_{\text{mix}}$ can play an important role in the anomalous $\pi^0\rightarrow\gamma\gamma$ process if the QCD axion is heavy enough; otherwise, this correction can always be safely neglected.


The remainder of this article is organized as follows. In Sec.~\ref{sec:ChiralLagrangian}, we elaborate on the tree-level chiral Lagrangian at LO and NLO and the anomalous WZW Lagrangian accounting for the neutral pseudoscalar meson decays. Then, in Sec.~\ref{sec:2rrLagrangianWidth}, we calculate the complete anomalous pion decay width with the contribution from the mixing with axion up to $\mathcal{O}(p^6)$. 
The discussion and conclusions are presented in  Sec.~\ref{sec:CONCLUSION}. Finally, in order to provide the reader with a clear and comprehensive discussion of the mixing issue, we have included in Appendix \ref{App:diagonalization} a detailed procedure for the exact diagonalization of the axion-pion mixing up to NLO.

\section{Chiral Lagrangian with axion field}\label{sec:ChiralLagrangian}

After integrating
out the heavy
degrees of freedom 
in a path-integral formalism, 
the Lagrangian, including the axion field and the isospin-triplet pion field at LO in the chiral expansion, has the form
\begin{eqnarray} \label{eq:LO}
\mathcal{L}_{p^2}
=\frac{F_0^2}{4}\langle D_\mu U^{\dagger}D^\mu U\rangle+
\frac{F_0^2}{4}\langle \chi_a U^{\dagger}+\chi_a U^{\dagger}\rangle,
\end{eqnarray}
where the symbol $\langle...\rangle$ stands for taking trace in the flavor space, and $\chi_a$ takes the form 
\begin{eqnarray}  \label{eq:chia}
\chi_a=2 B_{0} \mathcal{M}_{q} \exp \left(i \mathcal{X}_{a} a/f_a\right), 
\end{eqnarray}
with $\mathcal{M}_q=\text{diag}\{m_u,m_d\}$ 
the diagonal quark mass matrix,
$F_0$ is the pion decay constant in the chiral limit, $U$ is a unitary SU(2) matrix incorporating the quantum fluctuations, and $B_0=-\langle\bar{q} q\rangle / F_0^2$ is related to the scalar quark condensate. 
In Eq.~(\ref{eq:LO}), the covariant derivative incorporating the external fields is given by 
\begin{eqnarray}
D_{\mu}=\partial_{\mu}-i\left[v_{\mu}, U\right]-i\left\{a_{\mu}, U\right\},
\end{eqnarray}
where $v_{\mu}$ and $a_\mu$ 
correspond to the external vectorial and axial currents, respectively. 

As mentioned in the introduction, 
CHPT is an effective theory 
of QCD in the low energy regimes, 
especially at low temperatures and low densities~\cite{Gorghetto-2018ocs,16Carignano.Mammarella.ea51503-51503PRD,Balkin-2020dsr}. In this work, we restrict ourselves to the two-flavor case. 
Thus, the degrees of freedom of the system are the isospin-triplet pion field and axion field. 
The former are the lightest mesons arising from the spontaneous chiral symmetry breaking, while the latter is introduced to the chiral Lagrangian according to the PQ mechanism. The unitary matrix  can be written as $U(x)=U_0\tilde{U}(x)$~\cite{Acharya-2015pya}, with $\tilde{U}(x)$ collecting the chiral fields parameterized in the exponential parametrization 
\begin{eqnarray}\label{Uparameter}
\tilde{U}(x)= e^{i\Pi(x)/F_0},~~~
\Pi(x)=\left(
  \begin{array}{cc}
    \pi_3 & \sqrt{2}\pi^+ \\
    \sqrt{2}\pi^- & -\pi_3 \\
  \end{array}
\right),
\end{eqnarray}
where the special unitary matrix $U_0$ represents the ground state and which, without loss of generality, can be taken to have the following diagonal form~\cite{Guo-2015oxa}  
\begin{eqnarray}\label{U0exp}
U_0=\left(
  \begin{array}{cc}
    e^{i\varphi} & 0 \\
    0 &  e^{-i\varphi} \\
  \end{array}
\right). 
\end{eqnarray}
Note that, in general, $\pi_3$ in Eq.~(\ref{Uparameter}) does not denote the physical neutral pion because there may be mixing between the pion and axion fields in the LO Lagrangian, as will be discussed below.

At NLO, only the tree-level term proportional to the coupling $l_7$ contributes to the axion-pion mixing. 
However, in order to provide an exact diagonalization of the axion-pion mixing in Appendix~\ref{App:diagonalization}, we present here all  tree-level terms
in the $\mathcal{O}(p^4)$ chiral Lagrangian, excluding the derivative terms, namely
\begin{align} \label{eq:l37h13}
 \mathcal{L}^{(4,\text{tree})}= &~ \frac{l_{3}}{16}\left\langle \chi_{a} U^{\dagger}+U\chi_{a}^{\dagger}\right\rangle^{2}
-\frac{l_7}{16}\left\langle\chi_{a} U^{\dagger}-U\chi_{a}^{\dagger}\right\rangle^2 \nonumber\\
& +\frac{h_1+h_3}{4}\langle\chi_{a}\chi_{a}^{\dagger} \rangle+\frac{h_1-h_3}{2}\text{Re}(\text{det}\chi_a).
\end{align}
In the above, the constants $l_7$ and $h_3$ are scale independent, while the constants $l_3$ and $h_1$ contain both ultraviolet finite and divergent parts, which is necessary to eliminate the divergences that appear at higher orders. The latter are
related to the renormalized ones by~\cite{84Gasser.Leutwyler142-142AP}
\begin{eqnarray}
l_{3}=l_{3}^{r}-\frac{\lambda}{2},~~~~~~
h_1=h_1^r+2\lambda, 
\end{eqnarray}
with $\lambda$ the divergence at the space-time dimension $d=4$ in dimensional regularization,
\begin{eqnarray}
\lambda=\frac{\mu^{d-4}}{16 \pi^{2}}\left\{\frac{1}{d-4}-\frac{1}{2}\left[\ln (4 \pi)+\Gamma^{\prime}(1)+1\right]\right\},
\end{eqnarray}
where $\mu$ is the scale in dimensional regularization.

As usual, the axion-photon coupling, $g_{a \gamma \gamma}$, is defined by the following Lagrangian 
\begin{eqnarray}
\mathcal{L}_{a \gamma \gamma}=
\frac{1}{8} g_{a \gamma \gamma}
\epsilon^{\mu \nu \rho \sigma} a \mathcal{F}_{\mu v} \mathcal{F}_{\rho \sigma}. 
\end{eqnarray}
We note that there is a freedom~\cite{Lu-2020rhp} of choosing the diagonal matrix $\mathcal{X}_{a}$ satisfying $\left\langle\mathcal{X}_{a}\right\rangle=1$. If it is chosen as $\mathcal{X}_{a}
=\mathcal{M}_{q}^{-1}/\langle \mathcal{M}_{q}^{-1} \rangle $~\cite{86Georgi.Kaplan.ea73-78PLB,87Kim1-177PR}, 
then $U(x)=\tilde{U}(x)=e^{i \Pi(x) / F_{0}}$, and there is no 
axion-pion mixing 
term in the LO chiral Lagrangian. 
In this case, by further including the contribution term of the ratio of the 
electromagnetic and the color anomaly $\mathcal{E}/\mathcal{C}$, the complete $\mathcal{O}\left(p^{4}\right)$ axion-photon coupling can be written as 
\begin{eqnarray} \label{eq:garr4full}
g_{a \gamma \gamma}=
\frac{\alpha_{\text{em}}}{2 \pi f_{a}}\left(\frac{\mathcal{E}}{\mathcal{C}}-\frac{2}{3} \frac{4+z}{z+1}\right),
\end{eqnarray}
where $\alpha_{\mathrm{em}}\simeq1/137$ is the fine structure constant of QED, and $z=m_u/m_d$. We emphasize that in Eq.~(\ref{eq:garr4full}), the first term is model-dependent since its value depends on the PQ charges assumed for quarks. In contrast, the second term is model-independent and arises from the minimal coupling to QCD at the non-perturbative level. Note that the same result for axion-pion coupling can also be obtained with other choices of $\mathcal{X}_a$. In that case, the axion-pion mixing at LO must be taken into account.

\section{Two-photon decays of the neutral pion}
\label{sec:2rrLagrangianWidth}

The new axion contribution to the two-photon decay of neutral pions can be obtained using the WZW Lagrangian with an external photon field.
The corresponding Lagrangian is given by~\cite{Kaymakcalan-1983qq,Meissner-1987ge,01Kaiser76010-76010PRD}
\begin{align} 
\mathcal{L}_{\mathrm{WZW}}^\text{em} =&
 -\frac{N_c}{32 \pi^2} \epsilon^{\mu \nu \rho \sigma}\Big\{
 \langle U^{\dagger} \hat{r}_\mu U \hat{l}_\nu+i \Sigma_\mu(U^{\dagger} \hat{r}_\nu U+\hat{l}_\nu)  \nonumber\\ 
& -\hat{r}_\mu \hat{l}_\nu\rangle \left\langle v_{\rho \sigma}\right\rangle 
+\frac{2}{3}\left\langle\Sigma_\mu \Sigma_\nu \Sigma_\rho\right\rangle\left\langle v_\sigma\right\rangle\Big\},  
\end{align}
where 
$Q=\mathrm{diag}\{2/3,-1/3\}$ 
denotes the usual diagonal quark charge matrix for the two-flavor case.
Our goal in this section is to compute the 
two-photon decay width of $\pi^0$ up 
to NLO. 
The chiral Lagrangian with a minimal set of terms in the anomalous-parity strong sector at $\mathcal{O}(p^6)$ has been given in Ref.~\cite{02Bijnens.Girlanda.ea539-544EPJC}. 
In this work, only the terms proportional to the coupling constants $c_3^W$, $c_7^W$, $c_8^W$, and $c_{11}^W$ are relevant to the two-photon decays, which are given by
\begin{align}
\begin{aligned}
\mathcal{L}^{(6)}_{\text{ano}}= &~ ic_3^W\epsilon^{\mu \nu \rho\sigma}\left\langle\chi_{-}f_{+\mu \nu} f_{+\rho\sigma}\right\rangle \\
& +ic_7^W\epsilon^{\mu \nu \rho\sigma}\left\langle f_{+\mu \nu}\right\rangle\left\langle f_{+\rho\sigma} \chi_{-}\right\rangle \\
& +ic_8^W\epsilon^{\mu \nu \rho\sigma}\left\langle f_{+\mu \nu}\right\rangle\left\langle f_{+\rho\sigma}\right\rangle\langle\chi_{-}\rangle \\
& 
+c_{11}^W \epsilon^{\mu \nu \rho \sigma}\left\langle f_{+\mu \nu}\right\rangle\left\langle f_{+\gamma \rho} h_{\gamma \sigma}\right\rangle .
\end{aligned}
\end{align}

\begin{figure*}
  \includegraphics[width=0.85\textwidth]{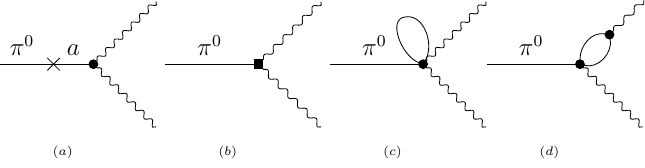}\\
  \caption{ The Feynman diagrams contribute to the anomalous decay of neutral pion: (a) axion-pion mass mixing, (b) $\mathcal{O}(p^6)$ tree-level Lagrangian, (c) the tadpole diagram from WZW Lagrangian, (c) the diagram with one vertex from the WZW Lagrangian and the other from $\mathcal{O}(p^2)$ Lagrangian, and (d) the diagram with both vertexes taken from $\mathcal{L}_{\text {WZW}}$. 
  }\label{fig:diagrams}
\end{figure*}

We have so far presented the chiral Lagrangian up to $\mathcal{O}\left(p^{6}\right)$, which is needed for our calculations, 
and we are now in a position to draw the Feynman diagrams responsible for the anomalous $\pi^0\rightarrow\gamma\gamma$ process and calculate the corresponding decay width, in particular the contribution from the axion-pion mixing at one-loop order. 
In Fig.~\ref{fig:diagrams}, we show the diagrams relevant for calculating the $\mathcal{O}\left(p^{6}\right)$ corrections to the decay width: (a) the axion-pion mixing from the NLO tree-level Lagrangian, (b) the tree-level diagram from $\mathcal{L}_{\text {ano }}^{(6)}$, (c) one-loop diagram with one vertex taken from $\mathcal{L}_{\text {WZW}}$ and the other one taken from the LO chiral Lagrangian, (d) one-loop diagram with both vertexes taken from $\mathcal{L}_{\text {WZW}}$.

As it is well known in the literature that the one-loop contributions from diagrams (c) and (d) are 
exactly cancel each other out~\cite{Donoghue-1985jkn}. Specifically, the upper photon line in diagram (d) is on shell, which results in the two contributions having opposite signs and ultimately canceling each other out. 
Compared to the previous studies, 
however, there is a non-vanishing new $\mathcal{O}(p^6)$ correction from the axion-pion mixing as shown in Fig.~\ref{fig:diagrams}(a), which is missing in all calculations of the decay width of $\pi^0\rightarrow\gamma\gamma$ process. 
Calling $\delta_{i}$ the relative correction at NLO to the amplitude for $\pi^0\rightarrow\gamma\gamma$ process, the complete decay width is given by
\begin{eqnarray}\label{eq:decaywidthFULL}
\Gamma_{\pi\gamma\gamma}=
\Gamma_{\pi\gamma\gamma}^{\text{(LO)}}
\left|1+\delta_{\text{tree}}+\delta_{\text{mix}}\right|^{2}, 
\end{eqnarray}
where the predicted LO decay width for $\pi^0\rightarrow\gamma\gamma$ process  
\begin{eqnarray}
\Gamma_{\pi\gamma\gamma}^{\text{(LO)}}
=\frac{\alpha_{\text{em}}^{2}}{(4 \pi)^{3}} \frac{m_{\pi}^{3}}{f_{\pi}^{2}}
=7.763 \pm 0.016~\text{eV},
\end{eqnarray}
which is about 5\% lower than the measurement from the PrimEx-II experiment at JLab, $\Gamma\left(\pi^0 \rightarrow \gamma \gamma\right)=7.802 \pm0.117$~\cite{PrimEx-II-2020jwd}. To resolve the discrepancy between the experimental data and the predicted LO decay width based on the chiral anomaly, the NLO corrections at $\mathcal{O}(p^6)$ are required. Nevertheless, it is important to stress that the estimates of the coupling constants, such as the combinations $5 \tilde{c}_{3}^{W}+\tilde{c}_{7}^{W}+2 \tilde{c}_{8}^{W}$ and $\tilde{c}_{11}^{W}/4+\tilde{c}_{3}^{W}+\tilde{c}_{7}^{W}$ in Eq.~(\ref{eq:deltaTREE}), are obtained by matching the two and three flavor couplings and using the experimental results for the decay rates of $\pi^0\rightarrow\gamma\gamma$ and $\eta\rightarrow\gamma\gamma$ processes as inputs~\cite{GrillidiCortona-2015jxo}. Consequently, we can not conversely use the numerical inputs, shown in Eq.~(\ref{eq:decaywidthFULL}), extracted from these decay rates of the neutral mesons to evaluate the decay width of $\pi^0\rightarrow\gamma\gamma$ process.

The correction $\delta_{\text{tree}}$ from the $\mathcal{O}(p^6)$ tree-level Lagrangian reads~\cite{Ananthanarayan-2002kj,Kampf-2009tk} 
\begin{align}\label{eq:deltaTREE}
\begin{aligned}
\delta_{\text{tree}} =  &~ \frac{16}{9} \frac{m_{\pi}^{2}}{f_{\pi}^{2}}\bigg[\frac{1-z}{1+z}\left(5 \tilde{c}_{3}^{W}+\tilde{c}_{7}^{W}+2 \tilde{c}_{8}^{W}\right)\\
& -3\left(\frac{\tilde{c}_{11}^{W}}{4}+\tilde{c}_{3}^{W}+\tilde{c}_{7}^{W}\right)\bigg],
\end{aligned}
\end{align}
and the correction $\delta_{\text{mix}}$ derived from the axion-pion mixing diagram is determined to be as follows
\begin{eqnarray}\label{eq:deltamix} 
\delta_{\text{mix}}
=2l_7\frac{m_a^2}{f_\pi^2(1+\beta_m)}\frac{1-z}{1+z}
\left(\frac{\mathcal{E}}{\mathcal{C}}-\frac{2}{3} \frac{4+z}{1+z}\right),
\end{eqnarray}
where the last part inside the brackets on the right-hand side of Eq.~(\ref{eq:deltamix}) comes from the axion-photon coupling constant $g_{a\gamma\gamma}$ at $\mathcal{O}(p^4)$ shown in Eq.~(\ref{eq:garr4full}). 
For convenience, we have defined $\tilde{c}_i^W \equiv\left(4 \pi f_\pi\right)^2 c_i^W$, and further replaced the factor $1/f_a^2$ with $m_a^2$ by using the relation between the axion mass and its decay constant, including the NLO correction derived in SU(2) CHPT~\cite{GrillidiCortona-2015jxo}, i.e.,
\begin{eqnarray}\label{eq:mafa}
m_a^2f_a^2=\frac{z}{(z+1)^2}m_\pi^2f_\pi^2(1+\beta_m),
\end{eqnarray}
where $\beta_m$ is given by
\begin{eqnarray}
\beta_m = 
2 \frac{m_\pi^2}{f_\pi^2}\bigg[h_1^r-h_3-l_4^r
+\frac{z^2-6 z+1}{\left(z+1\right)^2} l_7\bigg]
\end{eqnarray}
with the coupling constant $l_4$ replaced by the renormalized one by $l_4=l_4^r+2\lambda$. It is important to note that the replacement can always be done here since the QCD axion mass and its decay constant are uniquely related~\cite{Sikivie-2009qn,Chigusa-2023rrz}.

From Eq.~(\ref{eq:deltamix}), it is clear that the correction $\delta_{\text{mix}}$ tends to vanish as the axion-photon coupling constant approaches zero, due to the cancellation between the anomaly coefficient $\mathcal{E}/\mathcal{C}$ and the QCD correction to the axion-photon coupling constant in Eq.~(\ref{eq:garr4full}).
In addition, if the axion decay constant is large enough, particularly within the classical axion window $
10^9~\mathrm{GeV}\lesssim f_a\lesssim10^{12}~\mathrm{GeV}
$~\cite{Kim-2008hd,DiLuzio-2020wdo}, 
the correction $\delta_{\text{mix}}$ might be strongly suppressed. In this case, 
this contribution would not have any visible effect on the $\pi^0\rightarrow\gamma\gamma$ process.

We mention that, however, after revisiting the experimental constraints on QCD axions in
the MeV mass window, it has recently been argued that there is still a possibility for a viable QCD axion model with a mass in the
MeV range~\cite{Alves-2017avw}. Moreover, the so-called heavy QCD axion~\cite{Rubakov-1997vp,Berezhiani-2000gh,Hook-2014cda,Fukuda-2015ana,Albaid-2015axa} is a well-motivated extension of the standard QCD axion 
that address the quality problem~\cite{Kamionkowski-1992mf,Barr-1992qq,Ghigna-1992iv,Holman-1992us} while still solving the strong CP problem. The QCD axion can become heavy through various mechanisms, such as with the
help of the mirror copied sector of the Standard Model~\cite{Fukuda-2015ana,Hook-2019qoh,Dunsky-2023ucb} 
and with the inclusion of the small instanton effects~\cite{Flynn-1987rs,Choi-1988sy,Choi-1998ep,Gaillard-2018xgk,Gherghetta-2020keg}. 
Currently, major efforts such as the ArgoNeuT experiment at Fermilab are underway to search for the heavy QCD axion~\cite{Kelly-2020dda,Chakraborty-2021wda,ArgoNeuT-2022mrm,Co-2022bqq,Bertholet-2021hjl}. 
Without loss of generality, consider the addition of an extra mass term arising from other sources of the additional breaking of the PQ symmetry to the axial anomaly of QCD, like the hidden sector, to the effective axion Lagrangian~\cite{19Alonso-alvarez.Gavela.ea223-223EPJC,Bauer-2021wjo,Wang:2024tre} 
$\delta \mathcal{L}_a= M^2a^2/2$,  
the axion mass in Eq.~(\ref{eq:mafa}) can be rewritten as
\begin{eqnarray} \label{eq:maExtend}
&&m_a^2 \simeq M^2+\frac{m_\pi^2f_\pi^2}{f_a^2}\frac{z(1+\beta_m)}{(1+z)^2},
\end{eqnarray} 
where $M^2$ represents the contribution from the hidden sector. In such scenarios, the axion mass could potentially reach the MeV scale or even larger~\cite{Agrawal-2017ksf,Gherghetta-2016fhp,Dunsky-2023ucb}.

\begin{figure}[bt]
  \includegraphics[width=0.48\textwidth]{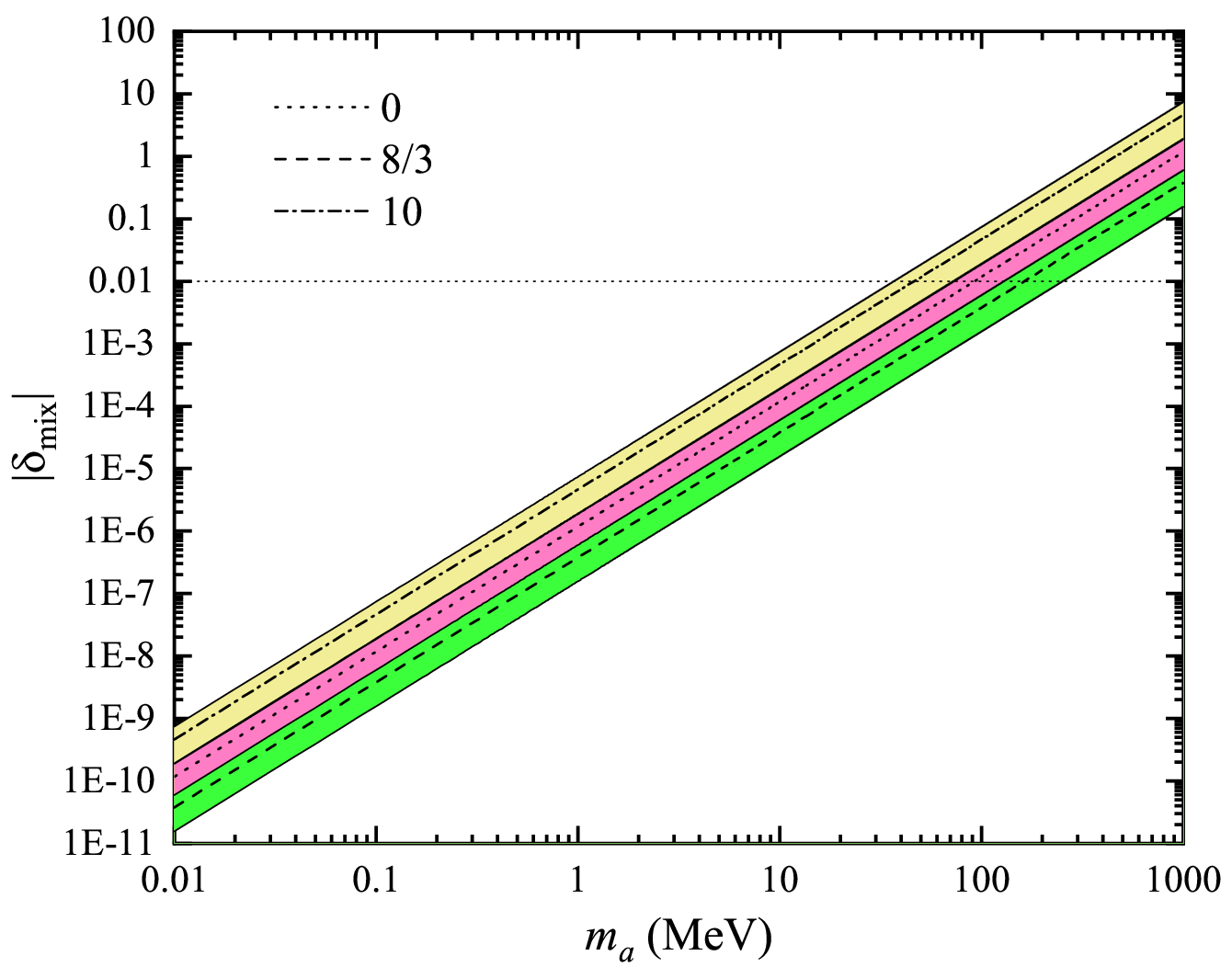}\\
  \caption{ The relative correction $\delta_{\text{mix}}$ to the pion decay width $\Gamma_{\pi\gamma\gamma}$ as a function of the QCD axion mass for several values of $\mathcal{E}/\mathcal{C}$. The dotted, dashed, and dash-dotted lines represent the relative correction $\delta_{\text{mix}}$ obtained with the central values of the parameters, while the shaded areas indicate the error bars. 
  }\label{fig:deltamixma}
\end{figure}

Assuming that the QCD axion is heavy enough and the relation between its mass and decay constant is well defined by Eq.~(\ref{eq:mafa}) in the considered energy scale, we plot the absolute value of the relative correction,  $|\delta_{\text{mix}}|$, 
as a function of the axion mass for several different representative values of $\mathcal{E}/\mathcal{C}$ 
in Fig.~\ref{fig:deltamixma}.  
From right to left, the dashed, dotted, and dash-dotted lines correspond to the results for $\delta_{\text{mix}}$ obtained with $\mathcal{E}/\mathcal{C}=8/3,~0,~10$ respectively, 
while the shaded area represents the error bars. 
Note that to perform the numerical calculations shown in this figure, we take the following values as inputs~\cite{GrillidiCortona-2015jxo}:
\begin{align}\label{eq:parameters}
\begin{aligned}
f_\pi & =92.2 ~\mathrm{MeV}, \\
m_\pi & =134.98 ~\mathrm{MeV}, \\
z & =0.48 \pm 0.03, \\
l_7 & =(7\pm 4) \times 10^{-3}, \\
h_1^r-h_3-l_4^r & =(4.8 \pm 1.4) \times 10^{-3} .
\end{aligned}
\end{align}

As can also be seen in Fig.~\ref{fig:deltamixma}, the relative correction $|\delta_{\text{mix}}|$ at low axion masses shows in all cases a similar behavior, that is, it increases monotonically with the increase of the axion mass. 
Consequently, from the results in Fig.~\ref{fig:deltamixma} we conclude that the relative correction $|\delta_{\text{mix}}|$ can always be safely neglected when the axion mass is smaller than the MeV scale. In particular, for the standard QCD axion with an eV or even lighter mass, the correction from the axion-pion mixing is always negligible. As mentioned above, however, we should emphasize that since the standard QCD axion suffers from the axion quality problem, axions with a much heavier mass and a lower decay constant have been proposed~\cite{Rubakov-1997vp,Berezhiani-2000gh}. 
In addition, there are certain mechanisms~\cite{Darme-2020gyx,17DiLuzio.Mescia.ea75003-75003PRD} that allow for the enhancement of the axion-photon coupling constant and thus the correction $\delta_{\text{mix}}$ without increasing the axion mass.
In these cases, the presence of an axion background might have visible effects on the anomalous decay process of $\pi^0\rightarrow\gamma\gamma$ if the axion mass is heavy enough.

We close this section by emphasizing that although the procedure for obtaining the analytical expression of the decay width of $\pi^0\rightarrow\gamma\gamma$ process in a QCD axion field background 
is formalized, the estimate of $\delta_{\text{mix}}$ up to the MeV mass range is just a qualitative 
illustration of the evolution of this correction with axion mass. 
Moreover, the plausibility
of this numerical estimate of $\delta_{\text{mix}}$ at the MeV mass scale would only be evident if one could obtain a convincingly precise 
formula for the QCD axion mass in this mass range.  
Instead, our main goal in this work is to show that there is a new contribution to the anomalous two-photon decay of the neutral pion in a QCD axion field background
at NLO, which is missing in all current calculations.  
Although this contribution can be safely neglected for the standard QCD axion with a mass smaller than the MeV scale, it is analytically unavoidable. On the other hand, the fact that the correction of $\delta_{\text{mix}}$ is extremely small might exclude the standard QCD axion as an explanation for the possible discrepancy between the CHPT prediction and the experimental data.

\section{Conclusions}  \label{sec:CONCLUSION}

As an effective field theory of the Standard Model at low energies, CHPT predicts mixing between the QCD axion and the neutral pion. 
We have studied the effect of a QCD axion field background on 
the decay width of a neutral pion into two photons 
in the framework of SU(2) meson CHPT, up to one-loop order. 
Our calculations indicate that an axion field background introduces an additional contribution to this anomalous decay process through axion-pion mixing. 
For the standard QCD axion, although the correction $\delta_{\text{mix}}$ is numerically negligible, it is analytically unavoidable. 
Remarkably, we have also discussed the possibility that if the axion mass could be as large as in the MeV or even larger mass range, such as the heavy QCD axion, the new contribution could potentially change the decay width of $\pi^0\rightarrow\gamma\gamma$ process, which could be comparable to the other contributions at one-loop order.

Our result provides insight into the behavior of the neutral pion in the presence of an axion field background. 
We mention that 
similar to the heavy QCD axion, ALPs 
can also couple to the neutral pion and have masses in the MeV range or beyond.  
In this sense, the discussion can also be extended to ALPs, although ALPs are not proposed to solve the strong CP problem and their masses are independent of the corresponding decay constant. 
We hope that this will stimulate work towards a more controlled determination of the coupling constants of CHPT 
and experimental searches for the axion, including the standard and heavy QCD axions and ALPs, as well as future research on the dynamics of the QCD axion and ALPs and their implications for cosmology, such as in a warm and dense medium~\cite{Zhang-2023lij,Gong:2024cwc}.

\section*{Acknowledgments}

We thank Mao-Jun Yan and Feng-Kun Guo 
for useful  discussions. ZYL acknowledges the warm hospitality of Institute of Theoretical Physics of Chinese Academy of Sciences, where part of this work was done.
This work is supported in part by 
the National Natural Science Foundation of
China (Grant Nos.~12205093, 12305096 and 12405054), and the Hunan Provincial Natural Science Foundation of China (Grant Nos.~2021JJ40188 and 2024JJ6210).

\appendix

\section{Exact diagonalization of the axion-pion mixing}
\label{App:diagonalization}

In the low energy regime, where the isospin-triplet pions and axions are the fundamental degrees of freedom, a mixing between the axion and the pion arises from the chiral Lagrangian, see e.g.~Refs.~\cite{Bauer-2017ris,Bauer-2020jbp,DiLuzio-2022tbb}. Although we can choose an appropriate $\mathcal{X}_{a}$ to avoid tree-level mixing at the LO between the axion and the pions, this mixing reappears in the $\mathcal{O}(p^4)$ order Lagrangian. In this appendix, we present a comprehensive diagonalization procedure for the axion-meson mixing up to the NLO.

By expanding the chiral Lagrangian in Eqs.~(\ref{eq:LO}) and (\ref{eq:l37h13}) up to the quadratic terms of the field and discarding the terms related to the charged meson field, we obtain
\begin{align}
\begin{aligned} \label{eq:quadra}
\mathcal{L}_{a\pi}=
&-\bigg[\frac{1}{2}m_\pi^2 
+\frac{l_3m_\pi^4}{f_\pi^2} 
-\frac{l_7m_\pi^4(1-z)^2}{f_\pi ^2(1+z)^2}\bigg]\pi_3^2 \\
&~ -\bigg[\frac{zf_\pi ^2m_\pi^2}{2f_a^2(1+z)^2} 
 +\frac{l_3zm_\pi^4}{f_a^2(1+z)^2} \\
&-\frac{4l_7z^2m_\pi^4}{f_a^2(1+z)^4} +{\frac{zm_\pi^4(h_1-h_3)}{f_a^2(1+z)^2} }\bigg]a^2 \\
&+\frac{4l_7zm_\pi^4(1-z)}{f_\pi f_a(1+z)^3}a\pi_3, 
\end{aligned}
\end{align}
where we have replaced $F_0$ by the physical pion decay constant $f_\pi$. 
From the last line of Eq.~(\ref{eq:quadra}), one 
notice that
the $\pi_3$ and $a$ states are mixed via an isospin violating term proportional to $a\pi_3$. 
The diagonalization of the $\pi_3, a$ mass matrix is given by the rotation
\begin{align}
\begin{aligned} \label{eq:rotation}
\binom{\pi^0}{a_{\text{phys}}} &= 
\left(
\begin{array}{cc}
\cos \varepsilon & \sin \varepsilon \\
-\sin \varepsilon & \cos \varepsilon
\end{array}
\right)
\binom{\pi_3}{a} \\
& =\binom{\pi_3+\varepsilon a}{-\varepsilon \pi_3+a}+\mathcal{O}\left(\varepsilon^2\right),
\end{aligned}
\end{align}
where $\pi^0$ and $a_\text{phys}$ denote the 
physical pion and axion fields respectively. 
Using Eq.~(\ref{eq:rotation}), we can derive expressions for $\pi_3$ and $a$ in terms of $\pi^0$ and $a_{\text{phys}}$, i.e.,
\begin{align}
\begin{aligned} \label{eq:pi3aexpre}
\pi_3 &=\pi^0-\varepsilon a_\text{phys}, \\
 a &=\varepsilon \pi^0 + a_\text{phys}. 
\end{aligned}
\end{align}
After substituting these expressions into Eq.~(\ref{eq:quadra}) and requiring that the coefficient of the term proportional to $a_\text{phys} \pi^0$ is zero, we obtain the analytical expression for the mixing angle, $\varepsilon$, between the axion and pion fields as 
\begin{align}
\begin{aligned} \label{eq:epsilon}
\varepsilon= &~ \frac{4l_7zm_\pi^4(z-1)}{f_af_\pi (z+1)^3} 
 \bigg[m_\pi^2 
 +\frac{2l_3m_\pi^4}{f_\pi ^2} \\
 &-\frac{2l_7m_\pi^4(z-1)^2}{f_\pi ^2(z+1)^2} \bigg]^{-1},
\end{aligned}
\end{align}
where we have dropped the higher order terms proportional to $1/f_a^2$. 
It is important to note that, in contrast to the axion-pion mixing scenario discussed in Ref.~\cite{DiLuzio:2024jip}, our analysis does not include any derivative mixing terms. This absence is due to our choice of a particular basis for $\mathcal{X}_{a}$, defined as $\mathcal{X}_{a} = \mathcal{M}_{q}^{-1}/\langle \mathcal{M}_{q}^{-1} \rangle$, which effectively precludes mass mixing between the axion and the neutral pion at LO. 
However, if we were to choose $\mathcal{X}_{a}=1/N_f$, where $N_f$ denotes the number of quark flavors considered, as in Ref.~\cite{Guo-2015oxa}, then we would have to consider the derivative mixing arising from the derivative terms in the chiral Lagrangian.

Having expressed the $\pi_3$ and $a$ states in terms of the physical pion and axion fields, as detailed in Eq.~(\ref{eq:pi3aexpre}), the correction term $\delta_\text{mix}$, shown in Eq.~(\ref{eq:deltaTREE}), can be easily derived from the merged effective Lagrangian $\mathcal{L}_{\mathrm{WZW}}^\text{em}+\mathcal{L}_{a \gamma \gamma}$. This correction analytically modifies the two-photon decay rates of the neutral pions.

\bibliography{RefLuInsp}

\end{document}